\newdimen\rotdimen
\def\vspec#1{\special{ps:#1}}
\def\rotstart#1{\vspec{gsave currentpoint currentpoint translate
   #1 neg exch neg exch translate}}
\def\rotfinish{\vspec{currentpoint grestore moveto}}
\def\rotr#1{\rotdimen=\ht#1\advance\rotdimen by\dp#1%
   \hbox to\rotdimen{\hskip\ht#1\vbox to\wd#1{\rotstart{90 rotate}%
   \box#1\vss}\hss}\rotfinish}
\def\rotl#1{\rotdimen=\ht#1\advance\rotdimen by\dp#1%
   \hbox to\rotdimen{\vbox to\wd#1{\vskip\wd#1\rotstart{270 rotate}%
   \box#1\vss}\hss}\rotfinish}%
\def\rotu#1{\rotdimen=\ht#1\advance\rotdimen by\dp#1%
   \hbox to\wd#1{\hskip\wd#1\vbox to\rotdimen{\vskip\rotdimen
   \rotstart{-1 dup scale}\box#1\vss}\hss}\rotfinish}%
\def\rotf#1{\hbox to\wd#1{\hskip\wd#1\rotstart{-1 1 scale}%
   \box#1\hss}\rotfinish}%

\documentstyle[12pt,epsfig,worldsci]{article}
\begin{document}
\def\z2{\ifmmode Z_2\else $Z_2$\fi}
\def\ie{{\it i.e.},}
\def\eg{{\it e.g.},}
\def\etal{{\it et. al.}}
\def\to{\rightarrow}
\def\tcv{\ifmmode t\to cV\else $t\to cV$\fi}
\def\Re{{\cal R \mskip-4mu \lower.1ex \hbox{\it e}\,}}
\def\Im{{\cal I \mskip-5mu \lower.1ex \hbox{\it m}\,}}
\setlength{\baselineskip}{2.6ex}

\vspace*{-1.3cm}
\begin{flushright}
ANL-HEP-CP-93-60 \\
OCIP/C-93-11\\
hep-ph 9309315
\end{flushright}
\vspace{-0.7cm}
\title{{\bf SUMMARY OF THE EXTENSIONS OF THE STANDARD MODEL
WORKING GROUP}}
\author{S.~GODFREY$^a$ and T.G.~RIZZO$^b$ \\
\vspace{0.3cm}
{\em $^a$Ottawa Carleton Institute for Physics, Department of Physics,
Carleton University,\\
Ottawa K1S 5B6, CANADA}\\
\vspace{0.3cm}
{\em $^b$High Energy Physics Division, Argonne National Laboratory,\\
Argonne, IL 60439, USA}\\}

\maketitle

\begin{center}
\parbox{13.0cm}
{\begin{center} ABSTRACT \end{center}
{\small\hspace*{0.3cm}
We summarize the results of the extended gauge group working group of
the Madison-Argonne Workshop on Present and Future Colliders.
Contributions are described on
the previously unexamined two photon fusion production of heavy
leptons, new studies of $Z'$ couplings to $\nu \bar{\nu}$ and
$q\bar{q}$, and previously unexplored vector leptoquark production.
More detailed accounts of these studies can be found in
individual contributions.}}
\end{center}

\section{Introduction}

A universal prediction of extensions of the standard model is the
existence of new particles.  Depending on the specific model under
consideration possible new particles may be extra gauge bosons, new types
of fermions, or leptoquarks.  The discovery of any of these new
particles  would dramatically show that the standard model is dead.  The
race would be on to solve the puzzle of the new standard model.
Thus,  there is a great deal of interest in constructing methods of
measuring the properties of newly found particles.
Ultimately, this knowledge
is necessary for understanding what the underlying theory might be.

In this context, the contributions to the {\it Extensions of the
Standard Model} working group consisted of an eclectic mix
of the previously unexamined two photon fusion production of heavy
leptons, new studies of $Z'$ couplings to $\nu \bar{\nu}$ and
$q\bar{q}$, and previously unexplored vector leptoquark production.
The analysis of the two photon fusion process found that it
is an important mechanism for the production of charged lepton pairs
providing the lepton mass, $m_L$, is below about 250 $GeV$.  The $Z'$
studies showed how detailed studies of $Z'$
production and decay can
provide additional information on $Z'$ couplings,
necessary for the
unravelling of the nature of a newly discovered $Z'$ bosons.
The leptoquark group performed a first calculation of {\it vector}
leptoquark production at hadron colliders via $gg$ fusion.  They find
cross sections which are substantially larger that those obtained
earlier for the scalar LQ case which has important implications on
limits and properties of vector leptoquarks.

The results of these studies are
briefly summarized below.  We encourage the interested reader to read the
more detailed individual contributions to these proceedings.

\section{New Probes of $Z'$ Couplings at the SSC and LHC}

Exploring the nature of newly discovered particles is significantly more
difficult at hadron supercolliders, such as the SSC and LHC, than at $e^+e^-$
colliders such as LEP, SLC, or NLC. If a new neutral gauge boson, $Z'$, were
to be discovered in $pp \to Z' \to l^+l^-$ channel, it would be extremely
important to learn as much as possible about the nature of its couplings in
order to
identify {\it {which}} $Z'$ of the many proposed in the literature (if any!)
has been found. During the last few years there has been a great amount of
work by many authors on this problem{\cite {R1}}. This work has shown that
$Z'$ identification {\it {may}} be possible for relatively light $Z'$'s in
the 1 TeV mass range although almost all of the individual proposals have
suffered from some sort of weakness, usually associated with statistics or
backgrounds from SM processes. Clearly, it is of some importance to have as
many tools available as possible to deal with this identification problem as
they may be applicable to other forms of new, non-SM physics as well as to
the $Z'$ situation itself.

There are a rather large number of contenders for a possible $Z'$, a few of
which are rather well known and we limit ourselves to those few extended
electroweak models(EEM) in the
discussion below: ($i$)the Left-Right Symmetric Model(LRM){\cite {R2}},
which has a single free parameter, $\kappa=g_R/g_L$, the ratio of the
$SU(2)_R$ and $SU(2)_L$ coupling constants, the Alternative Left-Right
Model(ALRM){\cite {R3}}, a $Z'$ with SM-like couplings(SSM), and the $E_6$-
inspired Effective Rank-5 Models(ER5M){\cite {R4}}

In these proceedings, Hewett and Rizzo{\cite {R5,R6}} have discussed three
techniques which may provide additional information on the nature of the $Z'$
coupling to $\nu \bar \nu$ and $q \bar q$. For each of them the major obstacle
to overcome is background from SM processes. In all cases it was assumed that
the $Z'$ had already been discovered and had its mass and width well
determined via the usual Drell-Yan production mechanism.

The first case{\cite {R5}} examined was $Z' \to jj$ which was shown
to be observable in
some models in the analysis performed by the ATLAS collaboration{\cite {R1}}
provided sufficient jet pair mass resolution is available and the $Z'$ is not
too heavy. The idea is straightforward: first one takes the full dijet sample
(after smearing with the detector resolution) with pair masses($M_{jj}$) in
the range 0.7-1.5 $M_{Z'}$ and applies strong rapidity and
$p_t$ cuts, $-1 < \eta <1$ and $p_t>0.2M_{Z'}$. Since most QCD induced events
are t-channel dominated, this substantially increases the ratio of signal($S$)
to background($B$). Since the width-to-mass ratio of $Z'$'s is generally small
($\leq 0.05$), and will in fact be known, clearly almost all of the $Z'$ events
will be in the `signal' region 0.9-1.1 $M_{Z'}$. Removing this region from the
data sample, the dijet mass distribution remaining events are fit by a 7th
order polynomial once an overall factor of $M_{jj}^{-5}$ is removed.
Increasing the order of the polynomial was not found to improve the
$\chi^2/d.o.f.$ of the fit. Extrapolating this fit into the signal region and
subtracting, one is left with a potential excess of $Z'$ events which is then
fitted to either a Gaussian or Breit-Wigner distribution. Summing the event
excess and scaling by the number of $Z' \to l^+l^-$ events in the discovery
channel reduces systematic errors and provides a handle on the ratio
$R=\Gamma(Z' \to jj)/\Gamma(Z' \to l^+l^-)$, which is quite sensitive to the
various couplings of the $Z'$.

While this procedure works rather well for a $M_{Z'}$=1 TeV at both the SSC and
LHC at design luminosities in a number of different models, the situation
was shown to be much more problematic (though not impossible) at higher masses
due to a loss in statistical power.

In a second analysis{\cite {R6}}, Hewett and Rizzo propose a method to get
some information on the ratio{\cite {R1}}
$r_{\nu \nu Z}=\Gamma(Z' \to Z \nu \bar \nu)/\Gamma(Z' \to l^+l^-)$, which
has been proposed as a probe of $Z'$ couplings, for $Z'$ masses in the 1 TeV
range. The SM background to this decay arises from the $pp \to 2Z$ process
with one $Z \to \nu \bar \nu$ decay. However, Hewett and Rizzo found that
this background could be quite precisely determined by measuring the $p_t$
distribution of the $Z$ in the corresponding $pp \to Zl^+l^-$ process and
then rescaling by the ratio of the $Z \to \nu \bar \nu$ to $Z \to l^+l^-$
branching fractions as determined by LEP, after suitably acceptance
corrections and rapidity cuts are applied. For a 1 TeV $Z'$, demanding
200 GeV $< p_t(Z) <500$ GeV yields for a reasonable $S/{\sqrt {B}}$ for
several extended models which then allows for a determination of
$r_{\nu \nu Z}$. Unfortunately, this method too fails for larger $Z'$ masses
due to a loss in statistics.

A last possibility is to examine monojet event excesses arising from $Z'j$
associated production where the $Z'$ decays to neutrinos. Here one actually
needs to consider the four processes
$pp \to Zj, Z'j \to l^+l^-j, \nu \bar \nu j$ subject to the cuts
$|\eta_j|<2.5$ and $p_t^j>200$ GeV for a 1 TeV $Z'$. From the number of SM $Z$
induced $l^+l^-j$ events, as determined by a dilepton mass reconstruction,
the anticipated number of SM $Z$ monojet events can be determined via the same
rescaled as in the $r_{\nu \nu Z}$ procedure above. Subtracting this result
from the total event sample leaves us with a potential excess induced by the
$Z'$. Scaling the excess by the observed number of $Z'j \to l^+l^-j$ events
determines the ratio
$R_{\nu}=\Gamma(Z' \to \nu \bar \nu)/\Gamma(Z' \to l^+l^-)$, which is then
shown to be quite sensitive to the fermionic couplings of the $Z'$ and, for
most models, is found to lie in the range $0 \leq R_{\nu} \leq 3$. As is the
case for the
other procedures, this method ceases to work for heavier $Z'$'s due to a
loss in statistics.


\section{Vector Leptoquark Production at the SSC and Tevatron}

The existence of leptoquarks($LQ$), objects carrying both lepton($L$) and
baryon($B$) numbers with either spin-0 or spin-1, is predicted in many
extended electroweak models which attempt to place quarks and leptons on an
equal footing{\cite {L1,L2}}. LQ may be searched for either indirectly
through their effects on low energy processes or by direct production at
various colliders. From LEP we only know that all types of LQ must be more
massive than 45 GeV while from HERA we obtain bounds which depend sensitively
on the unknown strength of the LQ Yukawa coupling to quarks and leptons.
Hadron colliders provide us with another tool with which to find LQ. Since
they can be pair produced by $gg$ or $q \bar q$ fusion, obtainable bounds
depend only upon whether the LQ is spin-0 or spin-1 and their branching
fraction into $l^{\pm}j$ or $\nu j$. Such searches have already been performed
by both the CDF{\cite {L3}} and D0 Collaborations{\cite {L4}} at the Tevatron
for the case of spin-0 LQ pair production, the cross section for which has been
known for some time{\cite {L5}}.

For these proceedings, Haber, Hewett, Pakvasa, Pomarol, and
Rizzo{\cite {L6}} have
calculated the pair production rate for spin-1, {\it {vector}}, LQ at the SSC
and Tevatron from the $gg$  fusion mechanisms. They find cross
sections which are substantially larger than those obtained earlier for the
scalar LQ case. At the Tevatron, this leads to significantly higher mass
limits arising from the existing cross section bounds in comparison to scalar
LQ with the same $l^{\pm}j$ branching fraction. Unlike the spin-0 case,
however, the pair
production of vector LQ ($V$) involves an additional ambiguity arising from
whether or not the $V$ is assumed to be a {\it {gauge boson}}
which originates from some extended symmetry group.
For the gauge boson case, the various trilinear $gVV$ and quartic $ggVV$
couplings involving vector LQ and gluons are
completely determined. If, LQ are not gauge bosons a fair amount of freedom
exists in these respective couplings even when we demand CP conservation. Of
course, for the non-gauge case there is not much motivation to choose any
particular set of these couplings. Bl\" umlein and R\" uckl{\cite {L2}}
consider the case where the vector LQ are {\it {minimally}} coupled, \ie~they
have an ``anomalous magnetic moment" parameter, $\kappa=0$, whereas in the
gauge
boson case LQ must have $\kappa=1$. The parton-level $gg$ cross section in the
$\kappa=0$ non-gauge case is not expected to obey unitarity but to scale as
$\alpha_s^2 \hat s/M_V^4$ for large $\hat s$. For $\kappa=1$, however, this
same cross section behaves as $\alpha_s^2 /\hat s$ in the same limit. For
the $q \bar q$ process, which is subdominant at both Tevatron and SSC
energies, the $\kappa=1$ choice requires the existence of an additional
$s$-channel exchange to maintain unitarity just as both $\gamma$ and $Z$
exchanges must be included when obtaining the correct cross section for
$e^+e^- \to W^+W^-$. This tells us that in a gauge theory of vector LQ, a
new spin-1, massive color octet particle must also exist, a scenario
realized in both the Abbott-Fahri model as well as $SU(5)${\cite {L1,L2}}.
As stated above, however, the subdominance of the $q \bar q$ process renders
the properties of this new particle academic since its influence on the LQ
pair cross section is insignificant and thus production via $gg$ fusion only
is considered.Once the value of $\kappa$ is chosen, the calculation of
the parton-level differential cross section is straightforward but
algebraically cumbersome.

Single production of LQ's at hadron colliders is also possible but is more
model dependent, making use of the {\it {a priori}} unknown $ql$ or $\bar ql$
Yukawa couplings. For scalar LQ's, it is well known that this single
production scenario can lead to a larger cross section than pair production
out to very large LQ masses $if$ the Yukawa coupling is of order
electromagnetic strength or greater. A similar result has been found to apply
in the case of vector LQ's. However, as these Yukawa couplings may turn out
to be quite small one should not rely solely on this mechanism to provide a
source of LQ's in hadronic collisions.

For details of this analysis, the reader should consult the individual
contribution of these authors.

\section{Heavy Charged Lepton Pair Production Through Photon Fusion at
Hadron Supercolliders}

Heavy leptons, both charged and neutral, are a feature of many models which
extend the particle content of the Standard Model.
The Drell-Yan
\cite{hl3,hl4}, gluon fusion \cite{hl4}, and gauge boson fusion
\cite{hl5} mechanisms for the production
of heavy charged leptons in hadron collisions have been investigated in the
past. In these proceedings Bhattacharya, Kalyniak and Peterson present
a preliminary study of the two photon production mechanism\cite{hl1},
which has been overlooked so far.
Both the inelastic process $p p \rightarrow \gamma \gamma X \rightarrow
L^+ L^-  X$ and the
elastic process, $p p \rightarrow \gamma \gamma p p \rightarrow L^+L^- p p$,
were considered in a Weizs\"{a}cker-Williams approximation.

The total cross sections for heavy charged lepton pair production in $p p$
collisions, for the elastic and inelastic  two photon fusion processes
along with the Drell-Yan and gluon fusion processes
are shown in Fig. 1 as a function of the charged lepton mass,
for the SSC center of mass  energy of 40 $TeV$.
These curves were obtained using the HMRS-Set B structure functions.
They assumed only three
generations of quarks, with the top quark mass set at 150 $GeV$.
The gluon fusion cross sections is
shown for a Higgs mass of 150 $GeV$.

\begin{figure}[t]
\vbox{
\begin{center}
\vspace{-4cm}
\hspace{0.0cm}
\setbox1=\vbox{\mbox{\epsfig{figure=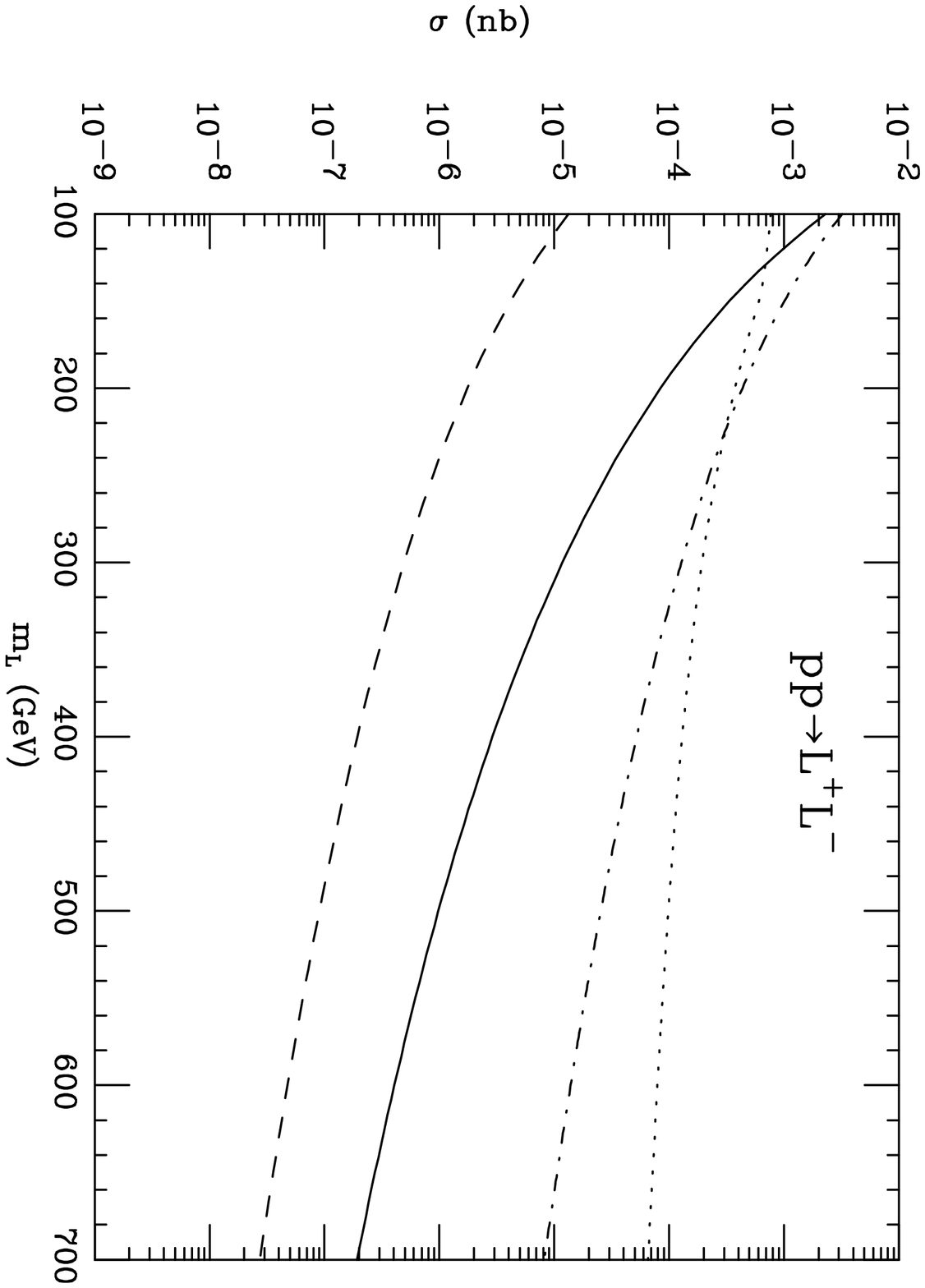,height=8cm}}}
\rotl{1}
\vspace{-4cm}
\end{center}
{\small Fig. 1 The total production cross section (in nanobarns) for a
charged lepton pair in $pp$ collisions at the SSC ($\sqrt s = 40$ TeV) as a
function of its mass $m_L$. The solid curve represents the lepton pair
production through two photon fusion in the deep inelastic scattering region
of protons, and the dashed curve shows the photon fusion production of
$L^+L^-$ for elastic collision of protons. The dotted and dot-dashed curves
represent respectively production through gluon fusion and the Drell-Yan
mechanisms.}
}
\end{figure}

The Drell-Yan process dominates for low
$m_L$. The gluon fusion cross section is
relatively flatter with increasing $m_L$ and
overtakes the Drell-Yan around $m_L$ of 240 $GeV$ for the SSC and
at about $m_L$ of 500 $GeV$ for the LHC.
The cross section for the inelastic two photon
process is within a factor of 1.4 of that for
the dominant Drell-Yan production when $m_L$ is 100 $GeV$.
The two photon inelastic production
falls to an order of magnitude below the now-dominant gluon fusion process
by $m_L$ of about 260 $GeV$.
Hence,  the process $p p \rightarrow \gamma \gamma X X
\rightarrow L^+ L^- X X$ is an important means of production of heavy charged
lepton pairs for
$m_L$ below 200-250 GeV at SSC energies which cannot be neglected in
studies of heavy lepton production at the SSC.
In contrast the inelastic two photon
process is much less important at LHC energies.


\section*{Acknowledgements}
We would like to take this opportunity to thank the collective members of
the LSGNA collaboration for insightful discussions during the course of the
Workshop.  S.G. thanks L.B. Hopson for helpful conversations.  This
work was partially funded by the Natural Sciences and Engineering
Research Council of Canada and by the U.S. Department of Energy.

\vspace{1.0cm}
%
\def\MPL #1 #2 #3 {Mod.~Phys.~Lett.~{\bf#1},\ #2 (#3)}
\def\NPB #1 #2 #3 {Nucl.~Phys.~{\bf#1},\ #2 (#3)}
\def\PLB #1 #2 #3 {Phys.~Lett.~{\bf#1},\ #2 (#3)}
\def\PR #1 #2 #3 {Phys.~Rep.~{\bf#1},\ #2 (#3)}
\def\PRD #1 #2 #3 {Phys.~Rev.~{\bf#1},\ #2 (#3)}
\def\PRL #1 #2 #3 {Phys.~Rev.~Lett.~{\bf#1},\ #2 (#3)}
\def\RMP #1 #2 #3 {Rev.~Mod.~Phys.~{\bf#1},\ #2 (#3)}
\def\ZP #1 #2 #3 {Z.~Phys.~{\bf#1},\ #2 (#3)}
\def\IJMP #1 #2 #3 {Int.~J.~Mod.~Phys.~{\bf#1},\ #2 (#3)}
\bibliographystyle{unsrt}

\end{document}